\documentclass[preprintnumbers, amsmath, amssymb,twocolumn, 
superscriptaddress, 
nofootinbib]{revtex4}
 \pdfoutput=1
 
\usepackage{graphicx}
\usepackage{epsfig}
\usepackage{bm}
\usepackage{amssymb}
\usepackage{float}
\usepackage{amsmath}
\usepackage{dcolumn}
\usepackage{cancel}
\usepackage[colorlinks]{hyperref}
\usepackage[usenames,dvipsnames]{color}
\hypersetup{
     breaklinks=true,
    pdfstartview={FitH},    
    colorlinks=true,       
    linkcolor=blue,          
    citecolor=red,        
    filecolor=magenta,      
    urlcolor=blue,           
    anchorcolor=green,      
    linktocpage=true
}

\def\doi{http://doi.org}

\begin{document}

\title{Do we need soft cosmology?}

\author{Emmanuel N. Saridakis}
\affiliation{National Observatory of Athens, Lofos Nymfon, 11852 Athens, 
Greece}
\affiliation{Department of Astronomy, School of Physical Sciences, 
University of Science and Technology of China, Hefei 230026, P.R. China}

\begin{abstract}  
 We examine the possibility of ``soft cosmology'',  namely small deviations 
from the usual framework due to the effective appearance of soft-matter 
properties in 
the Universe sectors. One effect of such a case would be the dark energy to 
exhibit a different equation-of-state parameter at large scales (which determine 
the universe  expansion) and at intermediate scales (which determine the 
sub-horizon
clustering and the large scale structure formation). Concerning soft dark 
matter, we show that it can effectively arise   due to the dark-energy 
clustering, even if dark energy is not soft. We propose a novel parametrization 
introducing the ``softness parameters'' of the dark sectors. 
As we see, although the background evolution remains unaffected,   due to the  
extreme sensitivity and significant effects   on the global properties  even a 
slightly non-trivial softness parameter  can improve  the clustering behavior 
and alleviate e.g. the $f\sigma_8$ tension. 
  Lastly, an extension of the cosmological perturbation theory and a detailed 
statistical mechanical analysis, in order 
to incorporate complexity and estimate the scale-dependent behavior  from first 
principles, is necessary and would provide a robust argumentation in favour of 
soft cosmology.

\end{abstract}


\maketitle

{\it Introduction} 

Standard cosmology has been proven very efficient, qualitatively and 
quantitatively, in describing the Universe evolution and properties at early 
and late times, as well as at large and small scales. Nevertheless, since 
cosmology has now become an accurate science, with the appearance of 
a huge amount of data of progressively increasing precision, slight 
disagreements, deviations and tensions between theory and observations lead to 
a large variety of extensions and modifications of the concordance paradigm.

Although in the usual ways of extensions one may add various novel 
fields, fluids, sectors and their mutual interactions \cite{Copeland:2006wr}, 
or alter the 
underlying gravitational theory \cite{Capozziello:2011et}, there is a  rather 
strong assumption that is  maintained in all of them, namely that the sectors 
that constitute the Universe are simple, or equivalently that one can apply the 
physics, the  hydrodynamics and thermodynamics of usual, ``hard''
matter. Nevertheless, in condensed matter physics it is well known that there 
is a large variety  of ``soft'' matter forms, which are characterized by 
 complexity,  simultaneous co-existence of phases, entropy 
dominance,  extreme sensitivity, viscoelasticity, etc, properties that arise 
effectively at intermediate scales  due to scale-dependent effective 
interactions  that are not present at the 
fundamental scales \cite{Jonesbook,Sagis2011}.

In this Letter we examine the possibility of ``soft cosmology'', namely 
small deviations from the usual framework due to the effective appearance of 
soft-matter properties in the Universe sectors. We mention   that due to the 
 extreme sensitivity and significant effects of softness on the global 
properties, one does not need to consider a large deviation from   standard 
considerations, since even a very slight departure would be adequate to improve 
the observed cosmological behavior at the required level. Finally, we stress 
that new fundamental physics is not directly needed, since the dark energy 
dynamical evolution and clustering, which is a widely accepted possibility in 
many scenarios beyond $\Lambda$CDM paradigm, is adequate to effectively induce 
the soft 
behavior.
 
 {\it Standard Cosmology} 
 
 Let us briefly review the basics of cosmology \cite{Peebles:1994xt}. The  
cosmological principle (the universe is homogeneous and isotropic at large 
scales) allows   to 
consider the Friedmann-Robertson-Walker (FRW) metric $ds^2= dt^2-a^2(t)\,
\delta_{ij} dx^i dx^j$.  
Concerning the universe content, one considers the usual 
baryonic matter and   radiation (i.e. all Standard Model particles), 
the dark matter sector, as well as the  
dark energy sector.  
The 
microphysics of   dark matter is unknown, and its source may be most probably 
some particle(s), however it may arise  from black holes, from 
modified gravity, or even from a combination of the above, i.e. a 
multi-component dark matter  
\cite{Bertone:2004pz}. The 
microphysics of   dark energy is unknown too, and it may arise from new fields 
or matter forms in the framework of general relativity, or it may have an 
effective nature of gravitational origin due to modifications of gravity.

The next step is to consider that (at least after a particular stage of the 
universe evolution) cosmological scales are suitably large in 
order to allow one to neglect the microphysics of the universe ingredients and 
describe them effectively through fluid dynamics  and continuum flow (at 
earlier stages one should use the   Boltzmann equation). Hence, 
one can ignore the microscopic Lagrangian of the various sectors, and write 
their energy momentum tensors as
$
 T^{(i)}_{\mu\nu}=(\rho_i+p_i)u_\mu u_\nu+p_i g_{\mu\nu},
$
with $\rho_i$ and $p_i$ the energy density and pressure of the fluid 
corresponding to the $i$-th sector, $u_\mu$   the 4-velocity vector field and 
$g_{\mu\nu}$ the metric. Note that one can extend the above expression by 
including viscosity or/and heat flux.

Under the above considerations, any cosmological scenario   will be determined 
by the two Friedmann equations
 \begin{eqnarray}
&&H^{2}=\frac{\kappa^2}{3}(\rho_b+\rho_r+\rho_{dm}+\rho_{de}),  \label{FR1} \\
&&  2\dot{H} + 3 H^2  = -\kappa^2 (p_b+p_r+p_{dm}+p_{de}),  
\label{FR2}
\end{eqnarray}
with $\kappa^2=8\pi G$, and where $H\equiv \dot{a}/a$ is the Hubble parameter. 
Additionally, the conservation equation 
$\nabla^\mu T_{\mu\nu}^{(tot)}=\nabla^\mu\left[\sum_i  T_{\mu\nu}^{(i)    
}\right]=0$ in the case of FRW geometry and for non-interacting fluids   gives 
rise to the separate conservation equations 
$  \dot{\rho}_{i} +3H  
(\rho_i+p_i )=0$,
while the extension to interacting cosmology can be   realized 
through  phenomenological   descriptors  $Q_i$  of the interaction
 with $\sum_i Q_i=0$ and with $\dot{\rho}_{i} +3H  
(\rho_i+p_i )=Q_i$.

In order for the equations to close we need to  impose the equation of state 
for each sector. The usual 
consideration is to 
assume barotropic fluids, in which the pressure is a function of the energy 
density only, with the simplest case being    $p_i=w_i\rho_i$ with  
$w_i$ the equation-of-state parameter.
Lastly, note that the above framework provides     $\Lambda$CDM 
cosmology for $\rho_{de}=-p_{de}=\Lambda/\kappa^2$, with $\Lambda$ the 
cosmological constant.
 
The above formulation of cosmological evolution allows one to proceed to a more 
subtle investigation, and study small perturbations around the FRW background. 
Focusing without loss of generality to the   linear theory of scalar 
isentropic   perturbations  in    the 
Newtonian gauge,   imposing $ds^2 =   -(1+2 \Psi)dt^2+      a^2 (t)(1-2 
\Phi) \gamma_{ij}  dx^i dx^j $, then in a general non-interacting scenario 
which includes the aforementioned sectors the scalar perturbations   are
determined by the equations \cite{Mukhanov:1990me,Ma:1995ey} 
\begin{eqnarray}
 &&\dot{\delta}_i+(1+{{w_i})\left(\frac{\theta_i}{a}-3\dot{\Psi}\right)+3H[c_{
\mathrm { eff } } ^ { (i)2 } -w_i ] %
\delta_i=0,\;}  \label{eq:line2} \\
 &&\dot{\theta}_i+H\left[1-3 c^{(i)2}_{\mathrm{ad}} 
\right]\theta_i-\frac{k^{2}c_{ \mathrm{eff}}^{(i)2}\delta_i}{(1+%
{{w_i})a}}-\frac{k^{2}\Psi }{a}=0,  \label{eq:line4}
\end{eqnarray}
 assuming zero anisotropic stress and with  $k$   the wavenumber of 
Fourier modes (in the    case of $\Lambda$CDM paradigm the  corresponding 
dark-energy perturbation equations are not considered). In the above equations 
  $\delta_i\equiv \delta\rho_i/\rho_i$ are the density
perturbations and $\theta_i$ is the divergence of the fluid velocity.
Furthermore,  $c_{\mathrm{eff}}^{(i)2}\equiv \delta p_i/\delta\rho_i$ and  
$c_{\mathrm{ad}}^{(i)2}\equiv w_i-\dot{w}_i/[3H(1+w_i)]$ are the effective and 
adiabatic sound
speed squares of the $i$-th sector respectively ($c_{\mathrm{eff}}^{(i)2}$ 
determines the amount of clustering). Note that the above equations can be 
simplified through the 
consideration of  the Poisson 
equation, which at sub-horizon scales can be written as 
\cite{Mukhanov:1990me,Ma:1995ey}:
$
-\frac{k^{2}}{a^{2}}\Psi =\frac{3}{2}H^{2}
\sum_i 
\left[ \left(1+3c_{\mathrm{eff}}^{(i)2}\right)\Omega_{i}\delta_{i}\right].
$ 
Finally, we note here that in general the above formulation can be  
applied also in the cases where the dark energy sector is an effective 
one arising from gravitational modifications.  
  
We close this section by mentioning that  one can find a big variety and many 
versions of the above formalism. However, there is a rather 
strong assumption that is  maintained in all of them, namely that the sectors 
that constitute the Universe are simple, or equivalently that one can apply the 
physics of usual
matter. In particular, the underlying assumption is that the laws that 
determine the Universe behavior at large scales can be induced by the laws that 
determine the interactions between its individual constituents.  Focusing on 
the hydrodynamic description, the use of fluid energy densities and 
pressures arises from the assumption that we can define fundamental
``particles'' of the corresponding sector, the collective flow of which gives 
rise to $\rho_i$ and $p_i$, while all physics below the particle scale has been 
integrated out.

 {\it Soft Cosmology} 
 
The above formulation of standard cosmology is definitely correct  and can 
provide a 
quite successful quantitative description of the Universe evolution. However, 
the question is: could it miss something on the details? Indeed, in principle 
there could be two sources of such information loss.

The first is that the integrated-out 
physics below the ``particle'' scale could leave different imprints on the 
physics above the  ``particle'' scale, depending on this coarse-graining scale. 
For instance
the correct 
integration out of the sub-galaxies physics leaves back a small but non-zero 
viscosity of the effective baryonic fluid \cite{Blas:2015tla}, while in 
all 
simulations of the galaxies and large-scale structure formation the assumptions 
on the integrated-out, subgrid physics are important, and the results are quite 
affected by it  \cite{Schaye:2014tpa}. This fact shows that 
one should use the coarse-grained description with caution, since using the 
same concept, of e.g the baryonic energy density $\rho_b$ and the effective 
hydrodynamic description, for scales that differ by   orders of magnitude 
is possible only after a correct integration-out of the neglected 
physics.

The second source of possible information loss is the assumption that the 
Universe sectors and interactions are simple and more or less 
scale-independent. 
For instance, two regions of the dark-matter fluid will mutual interact in the 
same way in the intermediate- and late-time universe, or the interaction of two 
big 
clusters of  dark matter   can arise by the superposition of all 
individual 
interactions of their sub-clusters, etc. In other words, in the concordance 
cosmological formulation one assumes that the sectors that constitute the 
Universe behave as usual, ``hard'' matter.

``Soft'' matter is a research field that has attracted a large amount of 
interest of the condensed matter community  
\cite{Jonesbook,Sagis2011}, since it has very 
interesting and peculiar properties far different than those of hard matter. 
The problem is that there is not a definition of what is soft matter. In 
particular, the best definition we have is that soft matter is the one that has 
the properties of soft materials. Examples of soft materials are the polymers 
(plastic, rubber, polystyrene, lubricants etc), the colloids (paints, milk, 
ice-cream etc), the surfactants, granular materials, liquid crystals, gels, 
biological matter (proteins, RNA, DNA, viruses, etc), etc. Although these 
examples of soft matter are very different from each other, they have some 
common properties and features that distinguish them from usual, hard, 
matter. Amongst others these include complexity (new qualitative properties 
arise at intermediate scales due to interactions that are not present at the 
fundamental scales), co-existence of phases (they have different phase 
properties depending on the scale at one examines them, e.g at the same 
time they can be fluid at small scales and solid at large scales), entropy 
dominance instead of energy dominance, flexibility, extreme sensitivity to 
reactions, viscoelasticity (they exhibit viscous and elastic properties 
simultaneously) etc.

In this work we examine the possibility that the dark sectors of the universe 
may exhibit (intrinsically or effectively) slight soft properties, which 
could then lead to small 
corrections 
to the corcodance model. We mention here that the discussion below holds 
independently of the underlying gravitational theory, i.e it is 
valid 
both in the framework of general relativity   as well as in modified 
gravity,  nevertheless  in the latter case we 
have richer possibilities to obtain scale-dependent 
interactions.

 {\it A. Soft Dark Energy} -- 
The nature and underlying physics of dark energy is unknown.
The basic framework that has been studied in extensive detail 
is 
that the dark energy fluid has the same fluid properties at all scales at a 
given moment/redshift.
However, as we 
saw, 
in soft matter the complexity that arises at intermediate scales may lead the 
material to have a different equation of state (EoS) at different scales 
simultaneously. 

As a simple phenomenological model of soft dark energy we may consider the 
case where dark energy has the usual EoS at large scales, namely at scales 
entering the Friedmann equations, but having a different value at intermediate 
scales, namely at scales entering the perturbation equations.  In this case 
the 
Universe's expansion history will remain identical to   standard cosmology, 
nevertheless the large-scale structure evolution can   deviate from the 
standard 
one and can be brought closer to observations. In summary one can obtain richer 
behavior. We mention that 
in the following we focus on sub-horizon scales $k\gg aH$ and thus to 
perturbation modes affected only by the intermediate-scale dark-energy EoS. The 
full analysis, in which different perturbation modes are affected by different 
EoS according to their scale, will be presented elsewhere.

A first approach on the 
subject would be to introduce the  effective ``softness parameter''  $s_{de}$ 
of the 
dark energy sector. This implies that while at cosmological, large 
scales (ls)   
dark energy  has the usual EoS, namely $w_{de-ls}$, at 
intermediate scales (is) we have
\begin{eqnarray}
\label{wdesoft}
&& w_{de-is}= s_{de}\cdot w_{de-ls},
\end{eqnarray}
and standard 
cosmology is recovered for $s_{de}=1$.
 
\begin{figure}[ht]
\includegraphics[width=6.7cm,height=4.4cm]{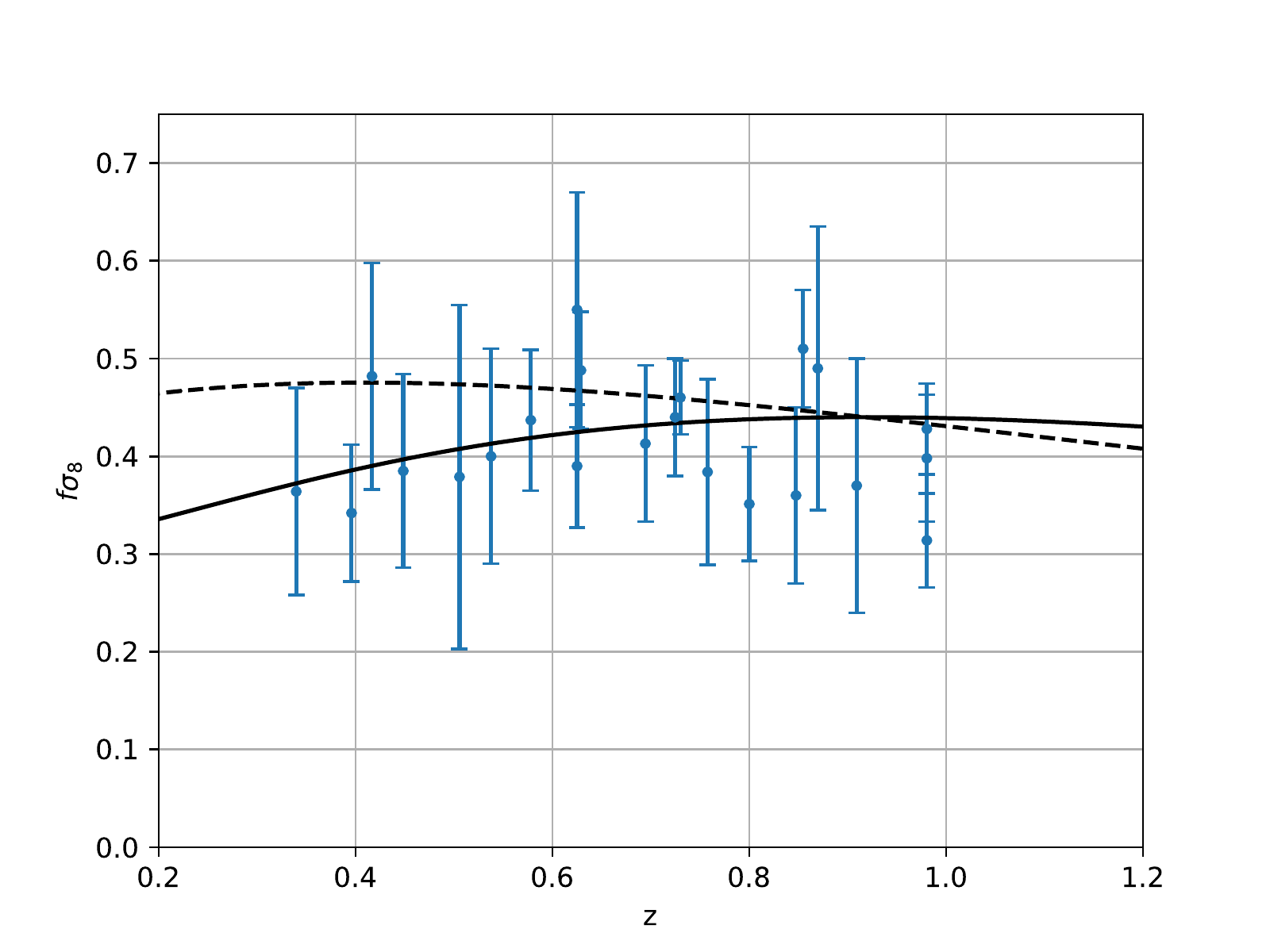}
	\caption{{\it{
	 The $f\sigma_8$ as a function of $z$. The dashed curve is for 
$\Lambda$CDM. 
The solid curve is for soft dark energy with $s_{de}=1.1$, i.e. with  
$w_{de-ls}=-1$ and
$w_{de-is}=-1.1$, 
and $c_{\mathrm{eff}}^{(de)}=0.1$, while dark matter is  standard  (i.e. 
not soft) with $w_{dm}=0$.
 }}}
	\label{fig1}
\end{figure}

For instance let us assume that the large-scale 
dark energy EoS 
$w_{de-ls}$ is a constant one $w_{de-ls}=w_0$ or e.g. the CPL one 
$w_{de-ls}=w_0+w_a(1-a)$. According to (\ref{wdesoft})  at intermediate 
scales the dark 
energy EoS $w_{de-is}$ is different, either constant $w_{de-is}=w_2$ or 
time-varying. Hence, the background Universe evolution will remain the same,
however 
since $c_{\mathrm{eff}}^{(de)}$ will change, through the Poisson 
equation and  
(\ref{eq:line2}),(\ref{eq:line4}) we will acquire a different evolution for the 
matter overdensity $\delta_m$. Hence, the resulting $f\sigma_8\equiv 
f(a)\sigma(a)$ value, with $f(a)=d\ln\delta_m(a)/d\ln a$ and 
$\sigma(a)=\sigma_8\delta_m(a)/\delta_m(1)$, will be different than the 
corresponding one of standard cosmology with the above  dark energy EoS (note 
that since the background behavior remains unaffected we do not need to worry 
about incorporating fiducial cosmology \cite{Aghanim:2018eyx}\footnote{The 
validity 
of secondary assumptions related to    $\sigma_8$ data formalism, such as the 
irrotational velocity field for the 
mater fluid \cite{Percival:2008sh}, should be carefully examined in the case of 
soft 
matter, nevertheless for small softness parameters, namely for small deviation 
from standard matter, one expects them to remain valid too.}). 
As we 
observe, we have a straightforward way to alleviate the $\sigma_8$ tension 
since we can suitably adjust $w_{de-is}$  in order to obtain slightly lower 
$f\sigma_8$. As a specific example    in Fig. \ref{fig1} 
we depict the $f\sigma_8$ as a function of $z$. The dashed curve is for 
$\Lambda$CDM. 
The solid curve is for soft dark energy with $s_{dm}=1.1$, i.e. with  
$w_{de-is}=-1.1$, 
and $c_{\mathrm{eff}}^{(de)}=0.1$, while dark matter is the standard one (i.e. 
not soft) with $w_{dm}=0$.
 Note that in principle $s_{de}$ can be varying too and one could introduce 
its parametrization, or one could additionally have more complicated situations 
in which  $ w_{de-is}$ and $ w_{de-ls}$ have  different parametrizations. In 
this first approach on soft dark energy we consider the simplest case of 
(\ref{wdesoft}).
 
Finally, note that soft materials may exhibit 
different EoS properties not only at different scales, but   at different 
directions too. In this case, one may think of a dark energy sector that has a 
different EoS at different directions, namely an anisotropic dark energy. 
However, such an analysis would require to deviate from FRW and consider 
explicitly anisotropic geometries such as the Bianchi ones. We will   
study this possibility in a separate work.

We close this paragraph by referring to  one of the properties  
of soft materials that can be relatively easily quantified, namely  
viscoelasticity \cite{Rheology2010}. In order to measure  
it one applies a 
sinusoidal strain $\epsilon=\epsilon_0\sin(\omega t)$ and measures the induced 
stress, which   for small $\epsilon_0$ is of the form 
$\sigma=\sigma_0\sin(\omega t+\delta)$. Pure elastic materials have $\delta=0$, 
 pure viscous materials have $\delta=\pi/2$, while viscoelastic materials have 
a non-trivial $\delta$ value. As a toy model to estimate the viscoelasticity of 
dark energy we start from the Friedmann equations (\ref{FR1}),(\ref{FR2}) and 
we perturb the scale factor solution $a(t)$ as   $a(t)\rightarrow 
a(t)+\epsilon_0\sin(\omega t)$, asking to see what will be the extra pressure 
(i.e. stress)   that one should obtain in (\ref{FR2}) in order to maintain 
consistency. One can easily see that for a general dark energy 
equation of state $w_{de}$ parametrization the resulting extra pressure will 
have a non-trivial $\delta$ (alternatively one could impose both a strain and a 
non-trivial extra pressure with a given $\delta$, and reconstruct suitably 
 $w_{de}(z)$ in order to obtain consistency, i.e. a dark-energy EoS 
parametrization with a desired viscoelasticity). 
From this simple and crude argument we deduce that a general dark energy has 
non-trivial 
viscoelasticity (we mention that viscoelasticity  
 is used as an extra argument, since by itself is not adequate to characterize 
a material as soft).

 {\it B. Soft Dark   Matter} -- 
In this subsection 
we examine the possibility that the dark matter sector exhibits soft 
properties. In the framework of general relativity    the gravitational 
interaction of dark matter with itself or with baryonic 
matter cannot produce internal complexity (unless the unknown microphysics of 
dark matter does impose an intrinsic soft structure). However, even if it 
is not intrinsic, soft behavior in the dark 
matter sector can still arise in an effective way due to the presence of 
non-trivial 
dark energy. Specifically, if the dark energy is clustering then, even if dark 
energy is not intrinsically soft, it will induce scale-dependent, qualitatively 
different 
intermediate structures in the dark matter clustering, at scales similar to the 
dark energy clusters. In particular, the interaction between two dark-matter 
clusters  below the dark-energy clustering scale (i.e. two dark-matter clusters 
with 
sparse dark energy between them) will be different from the interaction  
between two dark-matter clusters with a dark-energy cluster between them. 
Hence, one will have the effective appearance of screening effects at 
intermediate scales, and thus of complexity (this is the standard way that 
complexity appears in the colloids, namely due to the non-trivial, 
scale-dependent 
structure of the bulk between them). In summary, one could have a dark matter 
sector which at large scales, namely at scales 
entering the Friedmann equations,  behaves in the usual dust way, 
but which at 
intermediate 
scales, namely at scales entering the (sub-horizon) perturbation equations, it 
could slightly 
deviate from that.

\begin{figure}[ht]
\includegraphics[width=6.7cm,height=4.4cm]{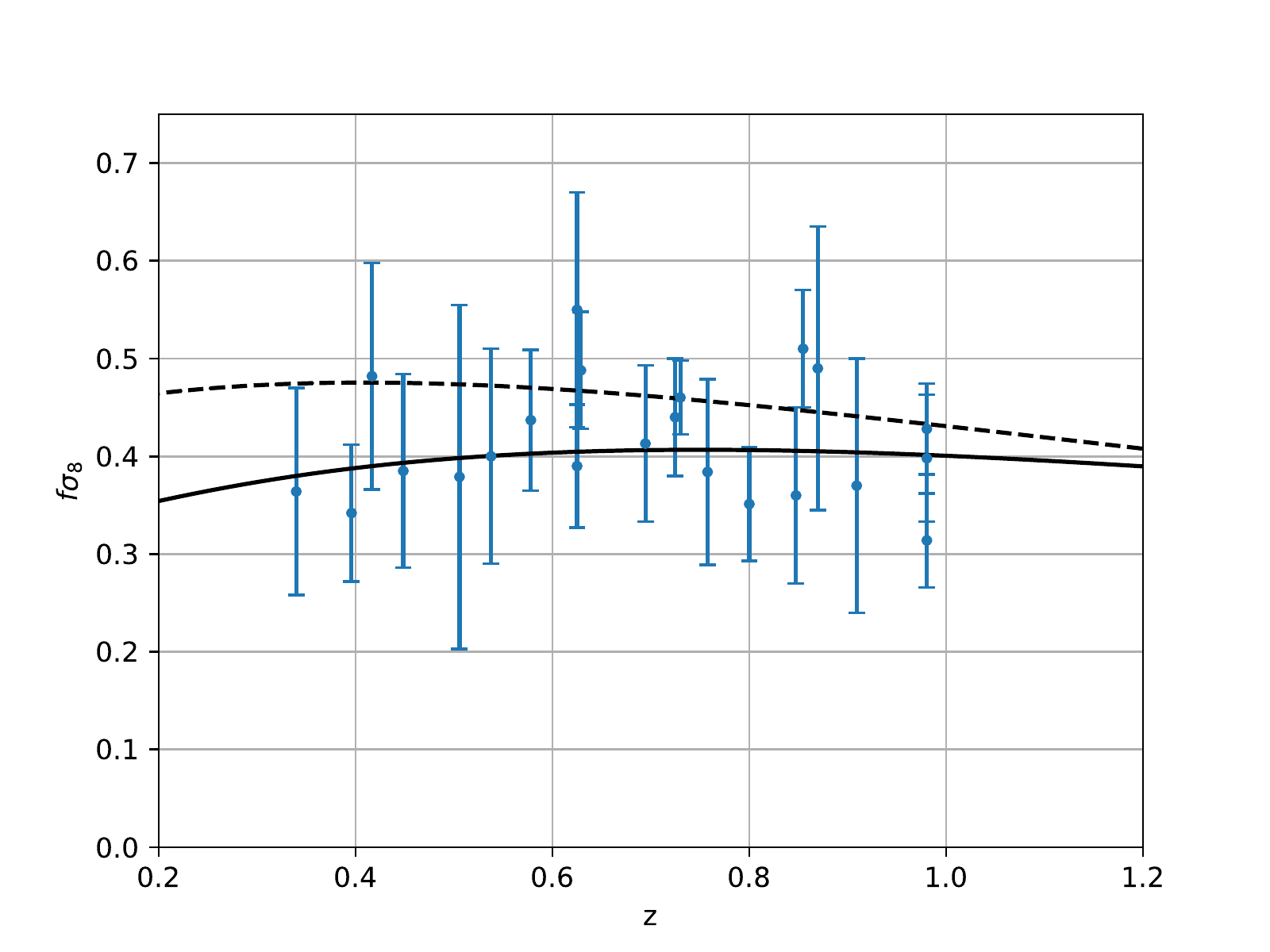}
	\caption{{\it{
	 The $f\sigma_8$ as a function of $z$. The dashed curve is for 
$\Lambda$CDM. 
The solid curve is for soft dark matter  with softness 
parameter $s_{dm}=1.05$, i.e. for dark matter  with $w_{dm-ls}=0$ and
$w_{dm-is}=0.05$    (note that dark energy is not soft). 
 }}}
	\label{fig2}
\end{figure}

We can introduce the dark matter   softness parameter $s_{dm}$  (standard 
cosmology is recovered for $s_{dm}=1$) as:
\begin{equation}
w_{dm-is}+1= s_{dm}\cdot (w_{dm-ls}+1),
  \label{wdmsoft}
\end{equation}
(mind the difference in the parametrization comparing to soft dark energy, in 
order to handle the fact that the dark 
matter EoS at large scales $w_{dm-ls}$ is 0).
Similarly to the example of the previous subsection, the 
background evolution will remain identical with that of standard cosmology, 
but the perturbation behavior (at sub-horizon scales) and the large-scale 
structure can be improved. 
In order to provide a specific example,  in Fig. \ref{fig2} 
we depict the $f\sigma_8$ as a function of $z$,
in the case of soft dark matter  with 
$w_{dm-is}=0.05$, i.e for dark matter  with softness 
parameter $s_{dm}=1.05$, while $w_{de-ls}=w_{de-is}=-1$.

In the above analysis we did not consider the dark energy sector to be soft. 
Definitely, proceeding to such a possibility would make the induced soft 
bahavior for dark matter easier. Moreover, this would be the case if one 
considers a mutual interaction between   dark matter and dark energy too, 
since a different than usual dark-energy clustering behavior would be 
transferred to a different than usual dark-matter clustering behavior, due to 
the interaction. Finally, deviating 
from general relativity would provide additional possibilities to induce 
effective soft properties to the dark matter sector, since dark matter will 
implicitly interact in a scale-dependent way (one would 
have the additional screening behavior due to the extra (scalar) graviton 
degrees of 
freedom that dress the dark matter in a scale-dependent way, altering  its 
self-interaction \cite{Joyce:2014kja}).

Let us make a comment here on the clustering features. The clustering behavior 
of soft matter has been extensively studied, and indeed it 
has been shown that the resulting  spectrum, factorial moments, fractal 
dimension, etc, depend on the specific intermediate-scale features. For 
instance 
the fractal dimension has been experimentally found to cover all the range from 
1 to 3 according to different materials (e.g. colloids of gold nanoparticle in 
aqueous media give $d_f=1.75\pm0.05$ for  diffusion-limited 
kinetics and  $d_f=2.01\pm0.10$ for  reaction-limited 
kinetics) \cite{PRL1985,Lazzari}).
On the other hand, the large scale structure and the galaxy distribution in the 
Universe has a fractal dimension $d_f=1.63 \pm 0.20$ 
\cite{Coleman:1992cm,Teles:2020aek}. 
The fact 
that soft matter clustering exhibits naturally non-trivial dynamics due to its 
intermediate-scale complexity, could be useful in describing the details of the 
observed large-scale structure. We mention here that the non-trivial clustering 
of soft matter changes at scales below the intermediate ones, and hence soft 
dark matter could alleviate the cuspy halo problem \cite{Teyssier:2012ie}, 
the dwarf galaxy problem \cite{Bullock:2017xww}, and other 
clustering-related problems that seem to puzzle the standard collisionless dark 
matter.


We close this subsection by referring to the possibility that the 
induced effective soft properties on the dark matter sector could lead to a 
direction-dependent EoS. Such slightly anisotropic dark matter might have a 
non-negligible effect on the lensing behavior. The detailed investigation of 
this possibility is left 
for a separate work. 

 {\it C. Soft Inflation} -- 
As a last application of soft cosmology we refer to the possibility that 
the inflation realization could exhibit soft features. Due to the 
extremely small length and time scales of the inflationary phase one may expect 
that complexity cannot be formed. Although this is indeed reasonable, due to 
the extreme sensitivity of the system behavior on even very small soft 
properties, one could still have the case that   slightly non-zero soft 
features could appear and then affect the inflation observables. For 
instance,  the inflation-driven field/fluid could develop a slight 
cluster structure during inflation,  inside the causal horizon, being either 
scale-dependent or direction-dependent. The   investigation 
of this possibility requires to deviate from the usual homogeneous and 
isotropic 
consideration.
Definitely, soft inflation could be thought as less possible than soft dark 
energy and soft dark matter, however it could still serve as the 
underlying 
mechanism of anisotropic inflation \cite{Watanabe:2010fh} and its role on 
explaining the 
possible non-trivial CMB anisotropies.

 {\it   Conclusions} 
 
We examined the possibility of ``soft cosmology'',  namely small deviations 
from 
the usual framework due to the effective appearance of 
soft properties in the Universe sectors. We started by considering   the 
possibility of soft   dark energy   due to 
intermediate-scale features that could arise  from its unknown microphysics. 
One effect of such a case would be the dark energy to exhibit a different 
EoS at large   and   intermediate scales. As we saw, 
although the background 
evolution remains unaffected, even a slight softness at intermediate scales can 
improve  the clustering behavior and alleviate e.g. the $f\sigma_8$ tension.

We proceeded to the examination of soft dark matter, which can effectively 
arise just due to the dark-energy clustering even if dark energy   is not 
soft. By considering a slightly different equation of state at 
large and intermediate scales we were able to improve the clustering behavior. 
Furthermore, in the additional incorporation  of soft dark energy, and/or 
modified gravity, the 
effective soft properties of dark matter could be richer, due to the extra 
screening mechanisms. Finally, for completeness we qualitatively presented the 
case of soft inflation.

We mention that in this work we incorporated   softness by 
phenomenologically introducing a slightly different EoS at different scales. 
Clearly, in order to incorporate complexity and estimate the scale-dependent 
behavior of the equation of states from first principles one should revise 
and extend  the cosmological perturbation theory and perform a detailed 
mesoscopic statistical mechanical analysis. Such a full investigation is 
necessary and would provide a robust argumentation in favour of soft cosmology 
(and towards this direction the Boltzmann-equation-based theoretical 
investigation 
of soft matter hydrodynamics will be used \cite{Marchetti2013}), nevertheless 
it lies beyond the 
scope of this initial investigation and will be performed elsewhere.
However, even in the present, phenomenological framework, since the 
rheology and dynamics of soft matter is different than the usual, hard one, one 
should confront it in detail with various observational datasets  and examine 
if   non-trivial softness parameters   are favoured (or even if they exhibit  
  more complex, scale-dependent or direction-dependent features). This 
detailed observational confrontation will be presented in a separate project.

 {\it Acknowledgments} --  
 The author wishes to thank P.J.E. Peebles for useful comments.

\end{document}